\documentclass[12pt]{article}
\usepackage[dvips]{graphics}
\usepackage{graphicx}
\title{Probability in relativistic quantum mechanics and
foliation of spacetime}
\author{Hrvoje Nikoli\'c \\
Theoretical Physics Division, Rudjer Bo\v{s}kovi\'{c} Institute, \\
P.O.B. 180, HR-10002 Zagreb, Croatia \\
{\normalsize e-mail: hrvoje@thphys.irb.hr} \\
\makebox[1in]{} \\
}
\date{\today}
\begin{document}
\maketitle
\begin{abstract}
The conserved probability densities (attributed to the conserved currents
derived from relativistic wave equations)
should be non-negative and the integral of them over an entire hypersurface
should be equal to one. To satisfy these requirements in a covariant 
manner, the foliation of spacetime must be such that each 
integral curve of the current crosses each hypersurface of the foliation
once and only once. In some cases, it is necessary to use 
hypersurfaces that are not spacelike everywhere.
The generalization to the many-particle case is also possible.
\end{abstract}
\vspace*{0.5cm}
\maketitle

\section{Introduction}

Finding a consistent probabilistic interpretation in the configuration 
space of relativistic quantum mechanics (QM) is a long-standing problem
(see, e.g., \cite{bjor1,schweber}). 
The simplest example leading to this problem
is the Klein-Gordon equation
(with the units $\hbar=c=1$ and the signature $(+,-,-,-)$)
\begin{equation}\label{KG}  
(\partial^{\mu}\partial_{\mu} +m^2)\psi(x)=0.
\end{equation}
The quantity $|\psi(x)|^2$ cannot be interpreted as the probability 
density because then the total probability $\int d^3x |\psi|^2$ 
would not be conserved in time. A better candidate for the 
probability density is the time component $j_0$ of the 
conserved current
\begin{equation}\label{cur}
j_{\mu}=i\psi^*\!\stackrel{\leftrightarrow\;}{\partial_{\mu}}\! \psi
\end{equation}
(where $a \!\stackrel{\leftrightarrow\;}{\partial_{\mu}}\! b \equiv
a\partial_{\mu}b - b\partial_{\mu}a$), 
but the problem is that $j_0$ may be negative on some regions of spacetime, 
even if $\psi$ is a superposition of positive-frequency plane waves only.
The Dirac equation of a single particle does not suffer from this 
problem, but a many-particle generalization of the Dirac equation leads to a 
similar problem \cite{bern}. 
The usual solution of the problem is second quantization (see, e.g.,
\cite{bjor2,schweber}), which postulates that $\psi$ is not a wave function 
determining probabilities, but an observable (called field) described by 
quantum field theory (QFT). 
Unfortunately, QFT is only a partial solution of the 
problem \cite{nikmyth}, 
because the axioms of QFT do not incorporate nor explain 
the probabilistic interpretation 
of $\psi$ in the nonrelativistic limit,
despite the fact that the probabilistic interpretation
of $\psi$ in the nonrelativistic limit is in agreement with 
experiments.
Finding a consistent relativistic position-operator could also solve the 
problem, but it seems that a hermitian position-operator cannot
be constructed in a covariant way \cite{newt,phil}.  
  
In this paper we propose a novel, Lorentz covariant, solution of the
problem of probabilistic interpretation of relativistic QM. 
The main technical ingredient is the {\em particle current}, 
which can be introduced either as a QFT operator 
\cite{wigh,nikolcur1,nikolcur2,nikolcur3}, or 
a c-number quantity calculated from the wave function attributed
to a QFT state \cite{nikolbohm1,nikolbohm2}. For simplicity,
in this paper we study 
free particles, but we note that the particle current can be introduced 
even when the interactions with classical or quantum fields (that cause 
particle creation and destruction) are present 
\cite{nikolcur1,nikolcur2,nikolcur3,nikolbohm1,nikolbohm2}.
The main conceptual ingredient is the observation that, 
despite common practice,    
there is no {\it a priori} reason why the hypersurface on which 
the probability is defined should be spacelike everywhere. 
Indeed, such hypersurfaces that are not spacelike everywhere may appear 
in some variants of the many-fingered time formulation of QFT 
\cite{oec,rov,dopl} and in the formulation
of QFT based on the covariant canonical De Donder-Weyl
formalism \cite{nikolepjc}. In this paper we show that hypersurfaces 
that are not spacelike everywhere naturally 
emerge from the requirement that the conserved c-number valued 
particle current should describe a probability density on a hypersurface.  

\section{Particle current}

For example, consider a hermitian (uncharged!)
scalar field operator $\hat{\phi}(x)$ 
that satisfies the Klein-Gordon equation (\ref{KG}). Denoting by 
$|0\rangle$ and $|1\rangle$ the Lorentz-invariant QFT states 
corresponding to the vacuum and a 1-particle state, respectively, 
the corresponding wave function \cite{schweber}  
\begin{equation}\label{wf}
\psi(x)=\langle 0|\hat{\phi}(x)|1\rangle
\end{equation}
is a superposition of positive-frequency plane waves only.
The corresponding c-number valued
particle current is given by (\ref{cur}) \cite{nikolbohm1}. Since 
$\psi$, just like $\hat{\phi}$, satisfies (\ref{KG}), it follows 
that the current is conserved:
\begin{equation}\label{cons}
\partial_{\mu}j^{\mu}=0 .
\end{equation} 

As another example, consider the electromagnetic field operator
$\hat{A}^{\alpha}(x)$ quantized using the covariant 
Gupta-Bleuler quantization \cite{schweber}. In this case, 
the 1-photon wave function is 
\begin{equation}
\psi^{\alpha}(x)=\langle 0|\hat{A}^{\alpha}(x)|1\rangle ,
\end{equation} 
while the particle current is
\begin{equation}\label{curphoton}
j_{\mu}=-i\psi^*_{\alpha}
\!\stackrel{\leftrightarrow\;}{\partial_{\mu}}\! \psi^{\alpha} .
\end{equation}
The photon wave function satisfies 
$\partial^{\mu}\partial_{\mu}\psi^{\alpha}=0$ and 
the current (\ref{curphoton}) 
satisfies the conservation equation (\ref{cons}).
For other examples of particle currents and their difference with respect to 
more familiar charge currents,
see \cite{nikolcur1,nikolcur2,nikolcur3,nikolbohm1,nikolbohm2}.

\section{Probability density and integral curves}

The currents defined as above have the property
\begin{equation}\label{globcons}
\int_{\Sigma} dS^{\mu}j_{\mu} =1,
\end{equation}
where $\Sigma$ is an arbitrary 3-dimensional spacelike hypersurface and
\begin{equation}
dS^{\mu}=d^3x |g^{(3)}|^{1/2} n^{\mu}
\end{equation}
is the covariant measure of the 3-volume on $\Sigma$.
Here $n^{\mu}$ is the unit future-oriented vector normal to $\Sigma$, 
while $g^{(3)}$ is the determinant of the induced metric on $\Sigma$.
The crucial consequence of (\ref{cons}) is that (\ref{globcons})
does not depend on the choice of the spacelike hypersurface $\Sigma$.
Owing to this fact, one is tempted to interpret the scalar density
\begin{equation}\label{j}
\tilde{j} \equiv |g^{(3)}|^{1/2} n^{\mu} j_{\mu}
\end{equation}
as the {\em probability density} $\tilde{p}$ on $\Sigma$. 
(The tilde above a quantity denotes that this quantity does not 
transform as a tensor, but rather as a tensor {\em density}.)
However, the probability density must satisfy the positivity 
requirement $\tilde{p}\geq 0$. Does $\tilde{j}$ satisfy 
the positivity requirement?

\begin{figure}[b]
\includegraphics[width=3.5cm]{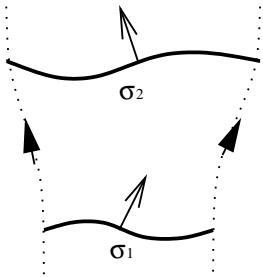}
\caption{\label{fig1}
Spacelike regions $\sigma_1$ and $\sigma_2$ with the property
$P_{\sigma_1}=P_{\sigma_2}$. The dotted curves are integral
curves of $j^{\mu}$, on which the arrows indicate the direction
of $j^{\mu}$. The arrows on $\sigma_1$ and $\sigma_2$
indicate the direction of the normal $n^{\mu}$.}
\end{figure}

First, consider the case in which $j^{\mu}$ is 
timelike and future-oriented everywhere. In this case, 
$\tilde{j}\geq 0$, so it is fully consistent to 
write $\tilde{p}=\tilde{j}$. The probability of finding 
the particle on some finite 3-dimensional region 
$\sigma\subset\Sigma$ is 
$P_{\sigma}=\int_{\sigma} d^3x\, \tilde{p}$. The probability 
$P_{\sigma}$ does not depend on $\sigma$ that
belongs to a family of $\sigma$'s constructed as follows 
(see Fig.~\ref{fig1}): The vector field $j^{\mu}(x)$ defines 
the congruence of integral curves, such that, at each point $x$, 
the vector $j^{\mu}$ is tangential to the curve. The integral curves 
$x^{\mu}(s)$ can be parametrized such that
\begin{equation}
\frac{d x^{\mu}}{ds}=j^{\mu} ,
\end{equation}
where $s$ is an affine parameter along the curve.
(For our purposes, these curves are only an auxiliary mathematical 
tool, see also \cite{lop} for the nonrelativistic case, but we 
note that in the Bohmian deterministic interpretation such 
integral curves represent actual particle trajectories 
\cite{bern,durr99,nikolbohm1,nikolbohm2,nikolbohmmeas}.) 
Consider the set of all integral curves that cross the boundary 
of some spacelike $\sigma_1$. This set of integral curves
defines a timelike hypersurface.
We say that a spacelike $\sigma_2$ 
belongs to the same family of $\sigma$'s as
$\sigma_1$ if the set of integral curves above 
crosses the boundary of $\sigma_2$
(see Fig.~\ref{fig1}).
The Gauss law and Eq.~(\ref{cons}) give
\begin{equation}\label{gauss}
\int_{\partial V^{(4)}} dS^{\mu}j_{\mu}=
\int_{V^{(4)}} d^4x \, \partial^{\mu}j_{\mu}=0 .
\end{equation}
Since $dS^{\mu}$ is orthogonal to $j^{\mu}$ on the timelike hypersurface 
defined by the integral curves, (\ref{gauss}) implies
$P_{\sigma_1}=P_{\sigma_2}$.  

\begin{figure}[b]
\includegraphics[width=7.4cm]{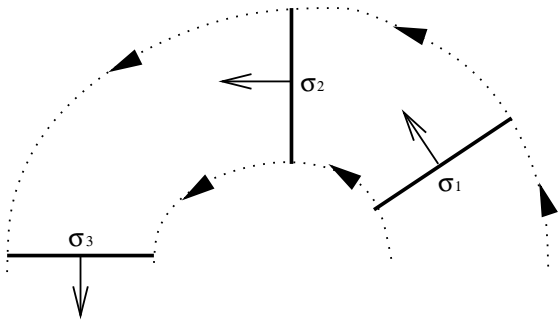}
\caption{\label{fig2}
Regions $\sigma_1$, $\sigma_2$, and $\sigma_3$
with the property $P_{\sigma_1}=P_{\sigma_2}=P_{\sigma_3}$.
Here $\sigma_1$ is null, $\sigma_2$ is timelike,
and $\sigma_3$ is spacelike but past-oriented.}
\end{figure}

The construction above is consistent for the case in which 
$j^{\mu}$ is timelike and future-oriented.
Indeed, if    
$\psi(x)$ is a plane wave $\propto\! e^{-ik_{\mu}x^{\mu}}$ with a positive
frequency $k^0=({\bf k}^2+m^2)^{1/2}$, then (\ref{cur}) is 
timelike and future-oriented. However, for a more general linear combination 
of plane waves with different positive frequencies, $j^{\mu}$ may not 
be timelike and future-oriented on some regions of spacetime. 
The generalization of the 
construction above to such a general case is the main aim of this paper.  
The basic idea is to consider hypersurfaces $\Sigma$
that may not be spacelike everywhere.
Instead, one can consider hypersurfaces
that are spacelike, null, or timelike on the regions
on which $j^{\mu}$ is timelike, null, or spacelike, respectively.
Can probability density be consistently defined on such 
regions? For that purpose, consider 
Fig.~\ref{fig2}, which represents an analog of Fig.~\ref{fig1}. 
On the spacelike but past-oriented region $\sigma_3$,
the timelike vectors $n^{\mu}$ and $j^{\mu}$ have the same direction,
so $\tilde{j}\geq 0$, which implies that $\tilde{p}=\tilde{j}$ 
on $\sigma_3$. On the timelike region $\sigma_2$  
the normal $n^{\mu}$ is spacelike, so the claim that 
$n^{\mu}$ is ``unit" actually 
means $n^{\mu}n_{\mu}=-1$. The vectors 
$n^{\mu}$ and $j^{\mu}$ have the same direction, which implies that
$\tilde{j}\leq 0$, so it is natural to take 
$\tilde{p}=-\tilde{j}=|\tilde{j}|$ on $\sigma_2$.
With these definitions of the probability density,   
the probability is conserved, i.e., $P_{\sigma_2}=P_{\sigma_3}$. 
This can be derived in the same way as for Fig.~\ref{fig1}, by using 
(\ref{gauss}) and the fact that 
$\sigma_2$ and $\sigma_3$ belong to the same family generated 
by the same set of integral curves of $j^{\mu}$.   
 
There remains one technical difficulty: how to define 
$\tilde{p}$ on a null region ($\sigma_1$ in Fig.~\ref{fig2})?
In particular, how to normalize the normal $n^{\mu}$, i.e., 
what does it mean that a null vector 
$n^{\mu}$ is ``unit"? To solve this problem, it is instructive 
to consider a simple example. Let $x^{\mu}$ be 
the standard orthogonal Lorentz coordinates
with the Minkowski metric. Consider also the coordinates 
\begin{equation}\label{coord}
x'^0=\frac{x^0-\beta x^1}{\sqrt{2}} , \;\;\;\;
x'^1=\frac{x^1+\beta x^0}{\sqrt{2}} ,
\end{equation}
where $\beta$ is a real constant. The coordinates $x'^0$ and $x'^1$
also represent two independent coordinate axes.
The hypersurface orthogonal to the axis
$x'^0$ is spacelike for 
$|\beta|<1$, null for $|\beta|=1$, and timelike for 
$|\beta|>1$. The normal to this hypersurface oriented in the direction 
of the axis $x'^0$ is 
\begin{equation}\label{nmu}
n^{\mu}=\frac{1}{\sqrt{|1-\beta^2|}} (1,-\beta,0,0) .
\end{equation}
Its norm is
\begin{equation}
n^{\mu}n_{\mu}={\rm sign}(1-\beta^2),
\end{equation} 
for $|\beta|\neq 1$. It is convenient to choose the coordinates 
$x'^1$, $x^2$, and $x^3$ as the coordinates on the hypersurface
orthogonal to the axis $x'^0$. 
From ({\ref{coord}), one finds 
$g'_{11}=-2(1-\beta^2)/(1+\beta^2)^2$, so the induced metric 
on this hypersurface has the property
\begin{equation}\label{g3}
|g^{(3)}|^{1/2}=\frac{\sqrt{2|1-\beta^2|}}{1+\beta^2} .
\end{equation}
We see that the case $|\beta|\rightarrow 1$ is singular. In particular, 
the components of (\ref{nmu}) become infinite, while the
quantity (\ref{g3}) becomes zero. However, this is only an apparent 
singularity, because the really relevant quantity in (\ref{j})  
is neither $|g^{(3)}|^{1/2}$ nor $n^{\mu}$, but rather their product
\begin{equation}
\tilde{n}^{\mu}=|g^{(3)}|^{1/2} n^{\mu}  .
\end{equation}
From (\ref{nmu}) and (\ref{g3}) we find
\begin{equation}
\tilde{n}^{\mu}=\frac{\sqrt{2}}{1+\beta^2} (1,-\beta,0,0) .
\end{equation}
This demonstrates the general rule that $\tilde{n}^{\mu}$ 
is well defined on all kinds of hypersurfaces, including 
the null ones. 

The results above can be summarized as follows.
For arbitrary $j^{\mu}$,
one considers hypersurfaces on which the normal $n^{\mu}$
is timelike, null, or spacelike on the
regions on which $j^{\mu}$ is timelike, null, or spacelike,
respectively.
%
%
On such hypersurfaces, the probability density is given by
\begin{equation}\label{density}
\tilde{p}=|\tilde{n}^{\mu}j_{\mu}| .
\end{equation}
%
%
However, even more general foliation of spacetime is admissible. 
The most general foliation 
that provides the conservation of probability is
{\em the foliation for which each
integral curve of $j^{\mu}$ crosses each hypersurface of the foliation
once and only once.}
Equations (\ref{globcons}), (\ref{gauss}), and (\ref{density})
imply that, on any such $\Sigma$,
\begin{equation}    
\int_{\Sigma} d^3x\, \tilde{p} =1 .
\end{equation}
In addition, for any $\sigma\subset\Sigma$, the probability
\begin{equation}
P_{\sigma}=\int_{\sigma} d^3x\, \tilde{p} 
\end{equation}
is invariant, i.e., does not depend on the choice of coordinates on
$\sigma$.
Note that, for $j^{\mu}$ as in Fig.~\ref{fig2}, it is impossible 
to find an admissible
foliation with hypersurfaces that are spacelike everywhere.
An example of an admissible foliation is sketched in Fig.~\ref{fig3}.

\begin{figure}[b]
\includegraphics[width=7cm]{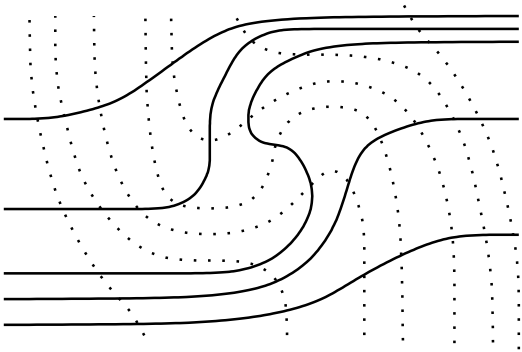}
\caption{\label{fig3}
A foliation (solid curves) of spacetime induced by the integral curves 
(dotted curves) of $j^{\mu}$.} 
\end{figure}

\section{Generalization to the many-particle case}

Let us also generalize the results above to many-particle states.
For example, the $n$-particle generalization of (\ref{wf}) is
\cite{schweber,nikolbohm1} 
\begin{equation}\label{wfn}
\psi(x_1,\ldots ,x_n)=(n!)^{-1/2}S_{\{ x_a\} }
\langle 0|\hat{\phi}(x_1)\cdots\hat{\phi}(x_n)|n\rangle ,
\end{equation}
where the symbol $S_{\{ x_a\} }$ ($a=1,\ldots ,n$)
denotes the symmetrization over all $x_a$, which is
needed because the field operators do not commute
for nonequal times. This $n$-particle wave function satisfies $n$ Klein-Gordon 
equations (\ref{KG}), one for each $x_a$.
The $n$-particle current generalizing (\ref{cur}) is
\begin{equation}\label{curn}
j_{\mu_1\ldots\mu_n}(x_1,\ldots,x_n)=i^n \psi^*
\!\stackrel{\leftrightarrow}{\partial}_{\mu_1}\! \cdots
\!\stackrel{\leftrightarrow}{\partial}_{\mu_n}\! \psi ,
\end{equation}
where $\partial_{\mu_a}\equiv \partial/\partial x^{\mu_a}_a$.
It transforms as an $n$-vector \cite{witt} and satisfies 
the conservation equation $\partial_{\mu_1}j^{\mu_1\ldots\mu_n}=0$ 
and similar conservation equations with other $\partial_{\mu_a}$.
The generalization of (\ref{globcons}) is
\begin{equation}\label{globconsn}
\int_{\Sigma_1} dS^{\mu_1}_1 \cdots \int_{\Sigma_n} dS^{\mu_n}_n \,
j_{\mu_1\ldots\mu_n} =1,
\end{equation}
which does not depend on the choice of timelike hypersurfaces
$\Sigma_1,\dots,\Sigma_n$. One can also introduce $n$ 1-particle currents 
$j_{\mu_a}(x_a)$ by omitting the integration over $dS^{\mu_a}_a$ in 
(\ref{globconsn}). For example, for $a=1$,
\begin{equation}
j_{\mu_1}(x_1)=
\int_{\Sigma_2} dS^{\mu_2}_2 \cdots \int_{\Sigma_n} dS^{\mu_n}_n \,
j_{\mu_1\ldots\mu_n}(x_1,\ldots,x_n) ,
\end{equation}
which does not depend on the choice of timelike hypersurfaces
$\Sigma_2,\dots,\Sigma_n$
and satisfies $\partial_{\mu_1} j^{\mu_1}=0$.
The wave function defined as in (\ref{wfn}) provides 
that different particles cannot be distinguished, which implies 
that $j^{\mu_a}(x)$ does not really depend on $a$. 
However, for a more general 
$n$-particle wave function, $j^{\mu_a}(x)$ may depend on $a$.
The integral curves of $j^{\mu_a}(x_a)$ determine
admissible foliations. The ``unit" normal on a hypersurface     
of such a foliation is $n^{\mu_a}(x_a)$, while  
the determinant of the induced metric on this hypersurface
is $g^{(3)}_a(x_a)$. Introducing
\begin{equation}
\tilde{n}^{\mu_a}(x_a)=|g^{(3)}_a(x_a)|^{1/2} n^{\mu_a}(x_a),
\end{equation}
the probability density generalizing (\ref{density}) is
\begin{equation}\label{densityn}
\tilde{p}(x_1,\ldots,x_n) = |\tilde{n}^{\mu_1}(x_1) \cdots
\tilde{n}^{\mu_n}(x_n) j_{\mu_1\ldots\mu_n}(x_1,\ldots,x_n)| .
\end{equation}
(Note that a construction similar to (\ref{densityn})
is discussed in \cite{durr99} for the case of fermions. 
However, in the case of fermions, it is not necessary to introduce
hypersurfaces that are not spacelike everywhere.)
The probability of finding one particle on $\sigma_1$, another 
particle on $\sigma_2$, etc., is
\begin{equation}
P_{\sigma_1,\ldots,\sigma_n}=\int_{\sigma_1} d^3x_1 \cdots
\int_{\sigma_n} d^3x_n \, \tilde{p}(x_1,\ldots,x_n) .
\end{equation}
This is a variant of the many-time probability $P(t_1,\ldots, t_n)$
\cite{tomon}, modified such that the regions $\sigma_a$ orthogonal to
$n^{\mu_a}$ may not be spacelike.

\section{Discussion and summary}

Now let us discuss the issue of causality related to 
the foliation of spacetime with hypersurfaces that are not
spacelike everywhere. 
One might think that such a foliation could be related to
particles that can move faster than light or backwards in time. 
Indeed, in the Bohmian deterministic hidden-variable interpretation,
such motions are possible \cite{durr99,nikolbohm1,nikolbohmmeas}.
However, the 
deterministic evolution of the wave function $\psi$ is, of course, causal, 
irrespective of the probabilistic interpretation of $\psi$. 
Consequently, with the conventional purely probabilistic interpretation 
on hypersurfaces that are not spacelike everywhere, one cannot use 
the foliation with
such hypersurfaces to send information to the past or faster than light. 
In this sense, {\em causality is not violated}.

Finally, let us make a few remarks on the problem of measurement.
We observe that, in nonrelativistic QM,
it is not so trivial to predict a probability density such as
$p(t,x,y)$ for a fixed $z$, despite the fact that
such a probability density can be determined experimentally.
The problem is that the corresponding measurements cannot be
attributed to only one equal-time hypersurface.
Consequently, to make predictions on such measurements,
one must deal with a theory of quantum
measurements that involves the problematic concept of
``wave-function collapse" or some substitute for it. Analogous problems
occur in our relativistic theory as well, when one wants to make
predictions on measurements that cannot be attributed to only one
admissible hypersurface. The solutions of such problems are expected
to be analogous to those in nonrelativistic QM, but a detailed
discussion of these aspects is beyond the scope of the present paper.

To summarize, in this paper we have shown that the conserved currents 
associated with relativistic wave equations can be consistently 
interpreted as probability currents. However, the main novel feature 
is that the shape of hypersurfaces on which the probability density 
can be defined depends on the direction of the current. In other words, 
the wave function (from which the current is calculated) determines 
not only the probability density, but also the admissible hypersurfaces
on which this probability density is defined. A curious but consistent 
feature of these hypersurfaces is that they may not be spacelike everywhere.

\section*{Acknowledgments}

The author is grateful to D.~D\"urr, S.~Goldstein, and R.~Tumulka 
for valuable discussions.
This work was supported by the Ministry of Science and Technology of the
Republic of Croatia. 

\end{document}